\documentclass[a4paper,fleqn,usenatbib]{mnras}

\usepackage{newtxtext,newtxmath}

\usepackage[T1]{fontenc}
\usepackage{ae,aecompl}

\usepackage{graphicx}	
\usepackage{amsmath}	
\usepackage{amssymb}	

\usepackage{slashed}

\title[Tests of $\Lambda$CDM and CG]{Tests of $\Lambda$CDM and Conformal Gravity using GRB and Quasars as Standard Candles out to $z\sim 8$}

\author[Carl Roberts, Keith Horne, Alistair O. Hodson, Alasdair Dorkenoo Leggat]{
Carl Roberts,$^{1}$\thanks{E-mail: cr214@st-andrews.ac.uk }
Keith Horne,$^{1}$ Alistair O. Hodson,$^{1}$ Alasdair Dorkenoo Leggat $^{1}$
\\
$^{1}$SUPA Physics and Astronomy, University of St Andrews, North Haugh, St Andrews, KY16 9SS, Scotland\\
}

\date{Accepted XXX. Received YYY; in original form ZZZ}

\pubyear{2017}

\begin{document}
\label{firstpage}
\pagerange{\pageref{firstpage}--\pageref{lastpage}}
\maketitle

\begin{abstract}
We compare the cosmology of conformal gravity (CG), \citep{Mannheim2006}, to $\Lambda$CDM. CG cosmology has repulsive matter and radiation on cosmological scales, while retaining attractive gravity at local scales. \citet{Mannheim2003} finds that CG agrees with $\Lambda$CDM for supernova data at redshifts $z<1$. We use GRBs and quasars as standard candles to contrast these models in the redshift range $0<z<8$. We find CG deviates significantly from $\Lambda$CDM at high redshift and that $\Lambda$CDM is favoured by the data with $\Delta\chi^2=48$. Mannheim's model has a bounded dark energy contribution, but we identify a $\lambda$ fine-tuning problem and a cosmic coincidence problem.
\end{abstract}




\section{Introduction}

Currently, $\Lambda$CDM is a highly successful theory, providing satisfactory agreement with supernovae data \citep{Riess1998}, the cosmic microwave background \citep{Planck2015} and galaxy clustering \citep{BOSS2017}, amongst others. However, dark matter and dark energy, required for these results, have their own issues \citep{Bullock2017}. There have been many indirect observations of dark matter, such as the Bullet Cluster \citep{Clowe2006}, but no confirmed direct detection. The cosmological constant has no firm theoretical basis, and the cosmological constant problem \citep{Weinberg1989} crops up when the amount of dark energy is derived according to the standard model of particle physics.

These problems motivate work on alternative gravity theories aiming to explain the observations without dark matter or dark energy, or to provide a new understanding for the cosmological constant. One such theory is conformal gravity (CG).  CG is an alternative gravity theory proposed by \citet{Weyl1918}, and more recently developed by \citet{Mannheim1989}. For a review of CG, see \citet{Mannheim2006}. CG has features that are attractive for a quantum gravity theory: renormalization, conformal symmetry, unitarity and no ghosts \citep{Bender2008}. 

In addition to the Newtonian potential, CG has linear and quadratic terms, with respect to radius, which are small compared to the Newtonian potential at solar system scales. Consequently, the solar system predictions are close to those of general relativity (GR). The linear and quadratic terms can be chosen to become important at galaxy scales, which allowed \citet{MannBrien2012} to fit the rotation curves of a sample of 111 spiral galaxies. However, the galaxy fits neglected the non-constant, non-zero Higgs field, whose conformal coupling is required to produce a nontrivial CG cosmology. The galaxy rotation curves of \citet{MannBrien2012} have a Higgs field that varies with radius, implying that particle masses change with radius. When  \citet{Horne2016} used a conformal transformation to make the Higgs field constant, there is an additional term from the Higgs field in the galaxy rotation curves, which nearly cancels out the linear term of the potential, leaving CG unable to reproduce the typical, flat rotation curves of galaxies. 

Because CG is a fourth-order theory, the scale of the quadratic potential (which is important for understanding galaxy rotation curves and lensing in CG) may be determined by the cosmology. Thus, it is important to understand the cosmological implications of CG, in addition to the galaxy scale effects. We look at the CG cosmology model, \citep{Mannheim1990}, who found no significant difference between CG and $\Lambda$CDM for supernova data for redshifts $z<1$. More recently, \citet{Diaferio2011} extended the Hubble diagram beyond supernovae (SNIa) by adding gamma-ray burst (GRB) standard candles to the supernova data to test the CG and $\Lambda$CDM cosmologies, and found that $\Lambda$CDM was slightly favoured. But when \citet{Diaferio2011} considered GRBs only, which probe higher redshifts than SNIa, $\Lambda$CDM was greatly preferred over CG.

In this paper, we analyse the \citet{Mannheim2006} cosmological model of CG, and compare its predictions to $\Lambda$CDM's using a Hubble diagram extended to redshift $z=8$ by including updated GRB and quasar standard candles. The outline for the paper is as follows: in Section 2, we briefly cover $\Lambda$CDM cosmology. In Section 3, we examine Mannheim's cosmology. In Section 4, we compare the predictions of the CG and $\Lambda$CDM with observations. We discuss various theoretical issues with the $\Lambda$CDM and Mannheim cosmologies in Section 5. Finally, we present our conclusions in Section 6.

\section{$\Lambda$CDM Refresher}

\begin{figure}
\begin{tabular}{c}
 \includegraphics[width=\columnwidth]{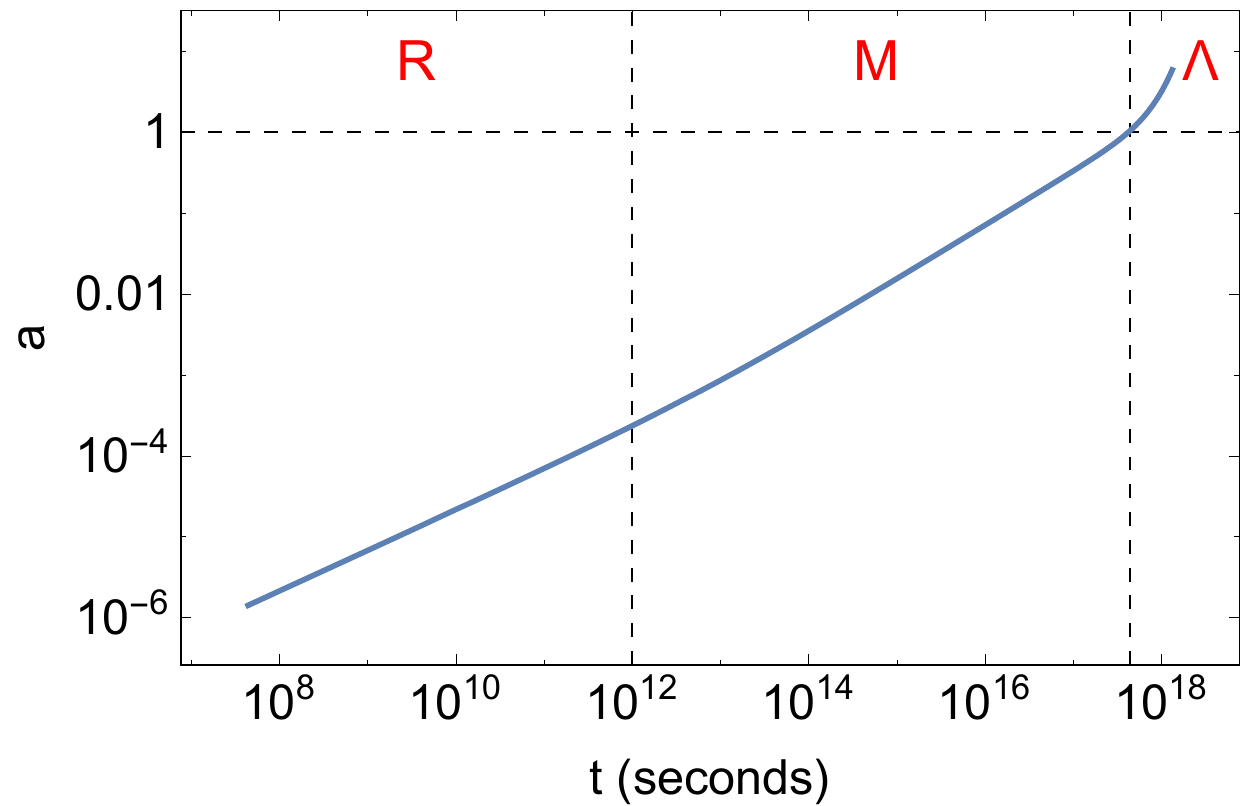}\\
 \includegraphics[width=\columnwidth]{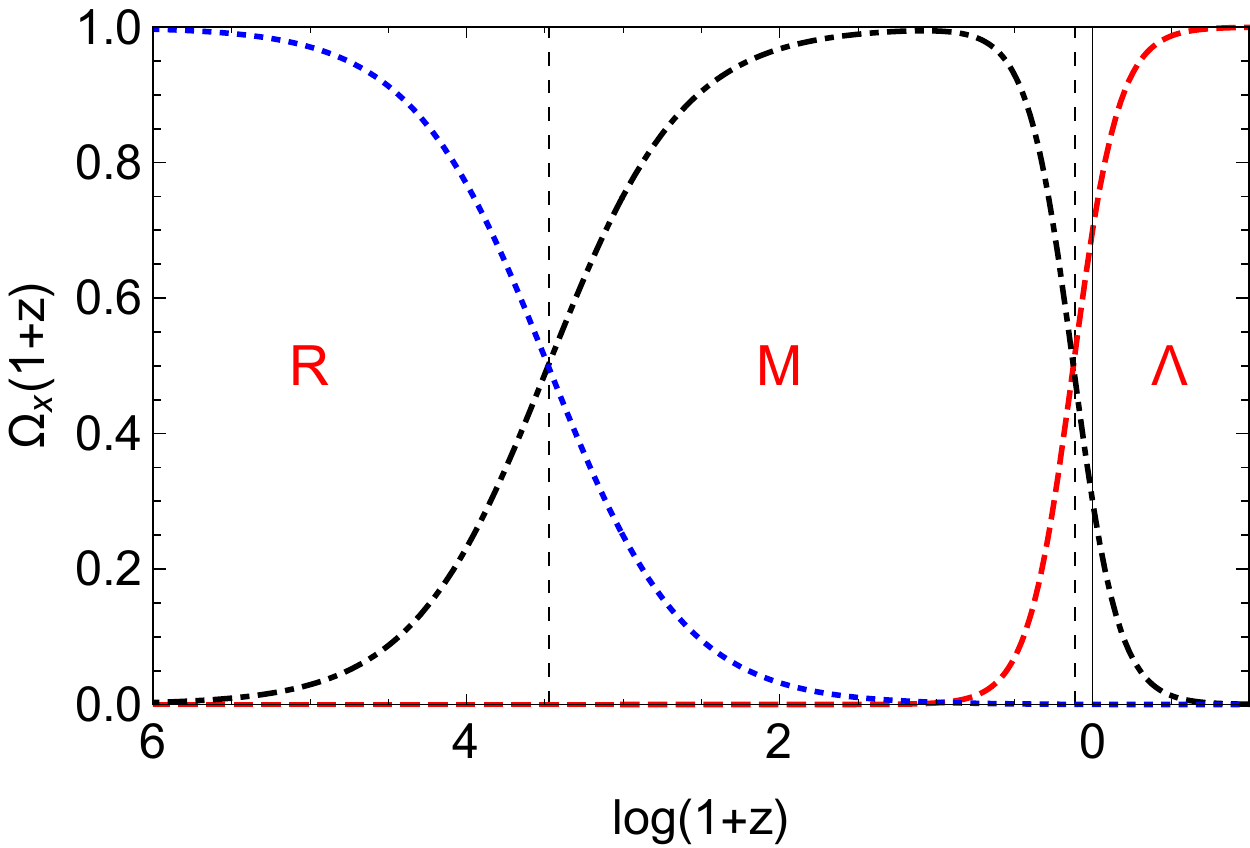}
\end{tabular}
 \caption{\emph{Top}: Scale factor vs time for $\Lambda$CDM, with $\Omega_{\Lambda}=0.7$ and $\Omega_{M}=0.3$. The majority of the past expansion occurs during the matter dominated Universe.
 \emph{Bottom}: The matter, radiation and dark energy contributions vs $\log(1+z)$ for the $\Lambda$CDM model. $\Omega_{\text{M}}(t)$ is the black dot-dashed line, $\Omega_{\Lambda}(t)$ is the red dashed line and $\Omega_{\text{R}}(t)$ is the blue dotted line. This model is transitioning from matter dominated to dark energy dominated at present time.}
 \label{fig:findark1}
\end{figure}

This paper follows the sign conventions of \citet{Mannheim2006}. The metric signature is ($-$,+,+,+), the Riemann tensor is $R^{\rho}_{\mu\sigma\nu}=-\partial_{\sigma}\Gamma^{\rho}_{\mu\nu}+...$, and the Ricci tensor is $R_{\mu\nu}=+R^{\sigma}_{\mu\sigma\nu}$. With these sign conventions, the Einstein-Hilbert action is:
\begin{equation}
    S_{\text{EH}}=-\frac{c^4}{16\pi G}\int d^4 x \sqrt{-g}\,(R-2\Lambda) \, ,
\end{equation}
where $R$ is the Ricci scalar and $\Lambda$ is the cosmological constant. Thus, the GR field equations are:
\begin{equation}
    -\frac{c^4}{8\pi G}G_{\mu\nu}=T_{\mu\nu}+\rho_{\Lambda}\, g_{\mu\nu}\, ,
\end{equation}
where $\rho_{\Lambda}=\frac{\Lambda c^{2}}{8\pi G} $, $G_{\mu\nu}$ is the Einstein tensor and $T_{\mu\nu}$ is the stress-energy tensor. The stress-energy tensor for a perfect fluid is: 
\begin{equation}
T_{\mu\nu}=(\rho+p)U_\mu U_\nu+p\,g_{\mu\nu}\, ,
\end{equation}
where $\rho$ is the energy density, $p$ is the pressure and $U_\mu$ is the perfect fluid 4-velocity, with $U_\mu U^\mu=-1$. Hereafter, we shall use natural units, $c=\hbar=1$. 

The homogeneous, isotropic spacetime is described by the Robertson-Walker (RW) metric:
\begin{equation}
 ds^2=-dt^2+a^{2}(t)\left( \frac{dr^2}{1-k\,r^2}+r^2 d\Omega^2 \right)\, ,    
\end{equation}
where $a(t)$ is the scale factor, $k$ is the curvature and $d\Omega^2=d\theta^2+\sin^2\theta\,d\phi^2$. $\Lambda$CDM uses a 3-component perfect fluid, each component having $p=w\rho$ with the equation of state parameter, $w=0,\,\frac{1}{3},\,-1$ for matter, radiation and dark energy respectively. The corresponding Friedmann equations (dropping the $t$ from $a(t)$ here on) of $\Lambda$CDM are:
\begin{align}
  \left( \frac{H}{H_0}\right)^2 &\equiv \left(\frac{\dot{a}}{a}\right)^2\,H_{0}^{-2} =\Omega_{\Lambda}+\Omega_K \,a^{-2} +\Omega_M \,a^{-3} +\Omega_R \,a^{-4}\, ,
  \label{LambdaEqn}
\\
  \Omega_K &=1-(\Omega_\Lambda +\Omega_M +\Omega_R)\, ,  
  \label{Eqn6}
\end{align}
where $\dot{a}$ denotes derivative with respect to time, $H_{0}$ is the current value of the Hubble parameter, $\Omega_{w}=\frac{\rho_w(t_0)}{\rho_c}$ with $\rho_c\equiv \frac{3H_{0}}{8\pi G}$ as the critical density at present time and $\Omega_{w}(t)=\frac{\Omega_{w}H_{0}^{2}}{a^{3(1+w)}H^{2}}$. Figure \ref{fig:findark1} serves as a useful comparison to later figures, showing how the scale factor evolves, beginning with a radiation-dominated phase $a\propto t^{1/2} $ that extends back to the singularity, through to a matter-dominated era $ a\propto t^{2/3}$, to finally a dark energy-dominated era $a\propto e^{H\,t}$. Throughout the rest of the paper, we shall refer to $\Omega_{M}=0.3,\, \Omega_{\Lambda}=0.7, \, \Omega_{R}=0$ as the $\Lambda$CDM model.

\section{Mannheim's Conformal Gravity Cosmology}

\begin{figure}
\begin{tabular}{c}
 \includegraphics[width=\columnwidth]{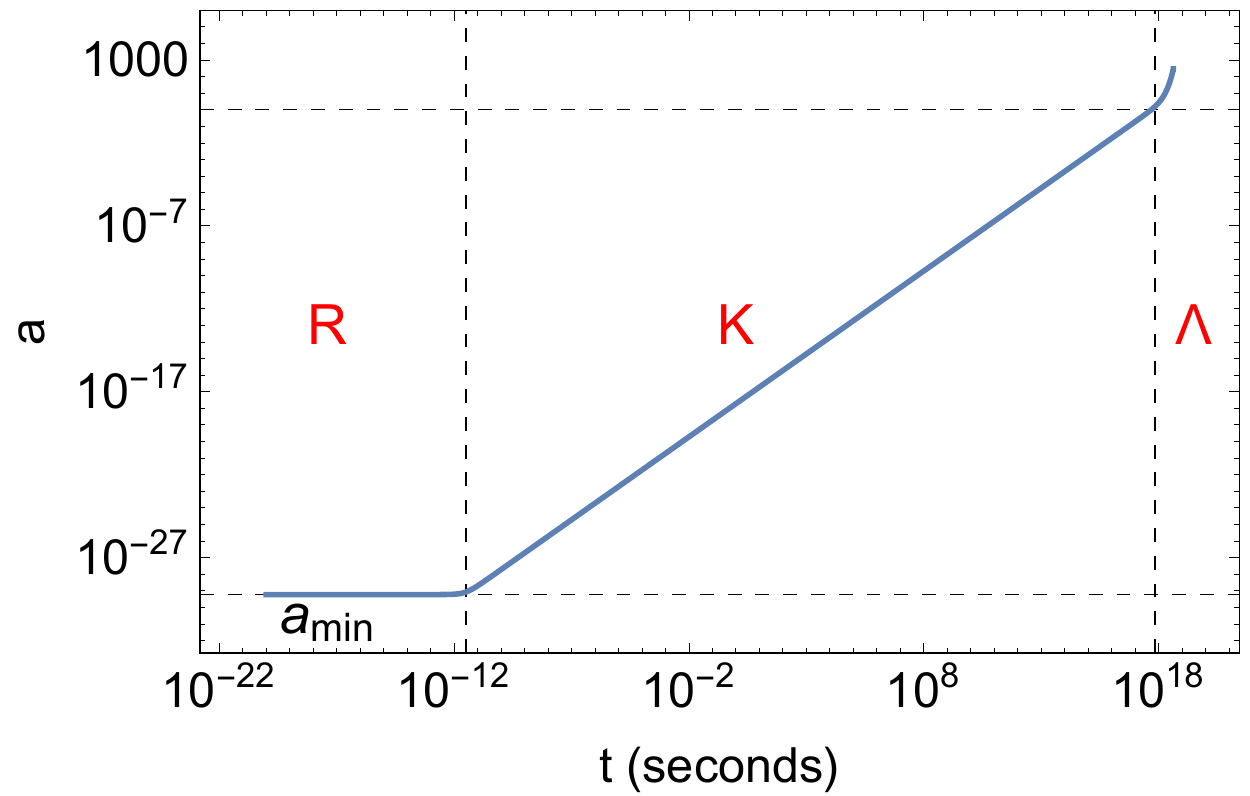}\\
 \includegraphics[width=\columnwidth]{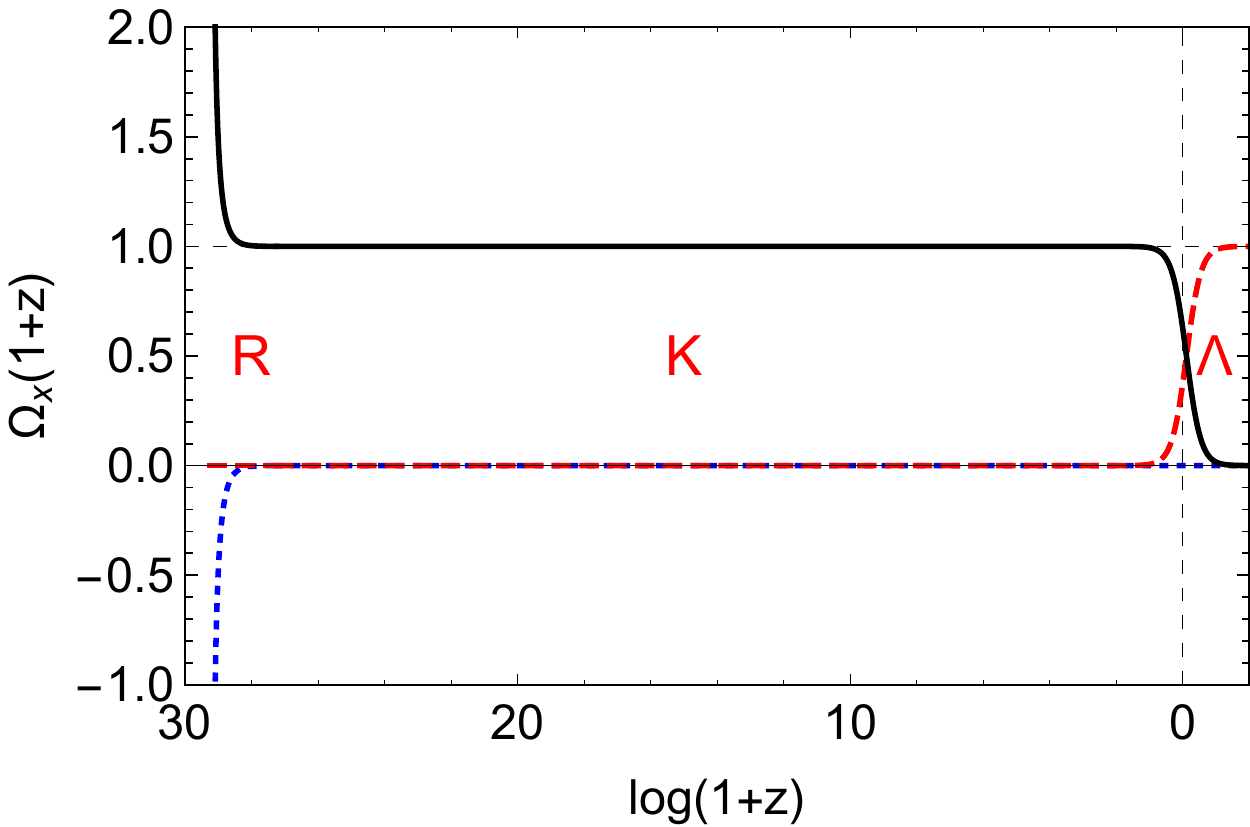} 
\end{tabular} 
\caption{\emph{Top}: Scale factor against time for Mannheim's model universe. The present scale factor is $a=1$. There is a minimum scale factor, $a_{\text{min}}=5.7\times10^{-30}$, so this model has no singularity at $t=0$. Thus, the maximum redshift is $z=1.75\times10^{29}$. The vertical dashed lines denote the time where the Universe transitions to curvature dominated and dark energy dominated respectively. The horizontal dashed lines are for $a=a_{\rm{min}}$ and $a=1$. 
\emph{Bottom}: The dark energy, curvature and radiation contributions plotted against $\log(1+z)$. $\Omega_{\text{K}}(t)$ is the black line, $\Omega_{\Lambda}(t)$ is the red dashed line and $\Omega_{\text{R}}(t)$ is the blue dotted line. There are three phases in this model: expansion from the minimum radius in the early Universe driven by radiation, a long period of linear growth via curvature, and an exponential era due to dark energy.}
\label{fig:MannheimOmega}
\end{figure}

Mannheim's CG cosmology is derived from the Weyl action instead of the Einstein-Hilbert action:
\begin{equation}
    S_{\text{Weyl}}=-\alpha_{g}\int d^{4}x\, \sqrt{-g} \, C_{\lambda\mu\nu\kappa}C^{\lambda\mu\nu\kappa}\, ,
\end{equation}
where $\alpha_g$ is a dimensionless constant and $C_{\lambda\mu\nu\kappa}$ is the Weyl or conformal tensor. CG is conformally invariant, whereas Einstein's GR is not. 

CG cosmology differs from $\Lambda$CDM by its matter action, which includes a conformal coupling of gravity to the Higgs field:
\begin{equation}
    S_{\text{M}}=-\int d^{4}x\, \sqrt{-g}\, \Big( \frac{1}{2}[S^{;\mu}S_{;\mu}-\frac{R}{6}S^{2}] +\lambda S^{4} +\bar{\psi}(\slashed{D}-m)\psi \Big)\ ,
    \label{Action}
\end{equation}
where $S$ is the Higgs field, $X^{;\mu}$ represents the covariant derivative, $\lambda$ is the dimensionless Higgs self-coupling constant, $\psi$ is the fermion's wave function, $\slashed{D}$ is the slashed Dirac operator and $m=hS$ is the fermion mass induced by coupling to the Higgs field.  In CG, the field equations analogous to the Einstein equations are the Bach equations:
\begin{equation}
4\alpha_g W_{\mu\nu}=T_{\mu\nu}\, ,
\end{equation}
where $W_{\mu\nu}$ is the Bach tensor, created by taking the variation of the Weyl action with respect to the metric, and $T_{\mu\nu}$ is the conformal stress-energy tensor. 

For the Robertson-Walker metric, the Bach tensor vanishes. Thus, all cosmological terms come from the stress-energy tensor, which is comprised of ordinary matter and radiation, treated as a perfect fluid, plus the scalar Higgs field. Working in the Higgs frame, where $S(t)=S_0$ and fermion masses are spacetime constants, we use a perfect fluid approximation for the Dirac terms, motivated by the incoherent averaging of \citet{Mannheim1990}. This leads to the stress-energy tensor taking on the form:
\begin{equation}
    0=T_{\mu\nu}=(\rho+p)U_\mu U_\nu+(p-\rho_{\Lambda})g_{\mu\nu}-\frac{1}{6}S_{0}^2 G_{\mu\nu} \, ,
\end{equation}
with $\rho_{\Lambda}=\lambda S_0^4$. The Higgs contribution to the matter action leads to a set of equations that in Mannheim's cosmology are identical to $\Lambda$CDM, except that instead of $G$, we have $-\epsilon G$, where $\epsilon=\frac{3}{4\pi G S_0^2}$:
\begin{align}
\left( \frac{H}{H_0}\right)^2 &=-\epsilon\left(\Omega_{\Lambda} +\Omega_M \,a^{-3} +\Omega_R \,a^{-4}\right)+\Omega_K \,a^{-2}\, ,
\label{MannEqn}
\\
\Omega_K &=1+\epsilon\left(\Omega_\Lambda +\Omega_M +\Omega_R\right)\, .
\label{Eqn13}
\end{align}

A full derivation from the action may be found in \citet{Mannheim2006} and Appendix A. Therefore, in Mannheim's model, homogeneous, isotropic matter and radiation are repulsive and the cosmological constant is derived from the Higgs vacuum energy, $\rho_\Lambda=\lambda S_{0}^{4}$. By defining $\bar{\Omega}_{w}\equiv-\epsilon\,\Omega_{w}$, we recover the same form as Eqn (\ref{LambdaEqn}) in $\Lambda$CDM but with $\bar{\Omega}_{\text{M}},\,\bar{\Omega}_{\text{R}}<0$ and, if we choose $\lambda<0$, $\bar{\Omega}_\Lambda>0$. Here on, we use this form of the equations and drop the bar notation.

Figures \ref{fig:findark1} and \ref{fig:MannheimOmega} show the evolution of the scale factor against time for both $\Lambda$CDM and Mannheim's model respectively. \citet{Mannheim2006} fits supernova data up to redshift $z\sim1$ using the model $\Omega_{\Lambda}=0.37, \, \Omega_{\text{K}}=0.63, \, \Omega_{\text{R}} \approx -10^{-60}, \, \Omega_{\text{M}}=0$. We refer to these values as the Mannheim model. \citet{Mannheim2006} requires $\Omega_{\text{R}} \approx -10^{-60}$ to have $T_{\rm{max}}> 10^{15}$K. One difference between the models is that $\Lambda$CDM has a singularity, whereas Mannheim's cosmology has a minimum scale factor, $a_{\rm{min}}=(\frac{-\Omega_{\text{R}}}{\Omega_{\Lambda}})^{\frac{1}{2}}=5.7\times10^{-30}$ and thus the maximum redshift is $z_{\text{max}}=1.75\times10^{29}$. In the very early Universe, radiation is dominant, and its repulsion prohibits a singularity at the origin. Therefore, a general feature of Mannheim's cosmology is that for models with non-zero matter and radiation contributions, the Universe has a maximum redshift, and thus a finite density, at $t=0$.

\begin{figure}
  \includegraphics[width=\columnwidth]{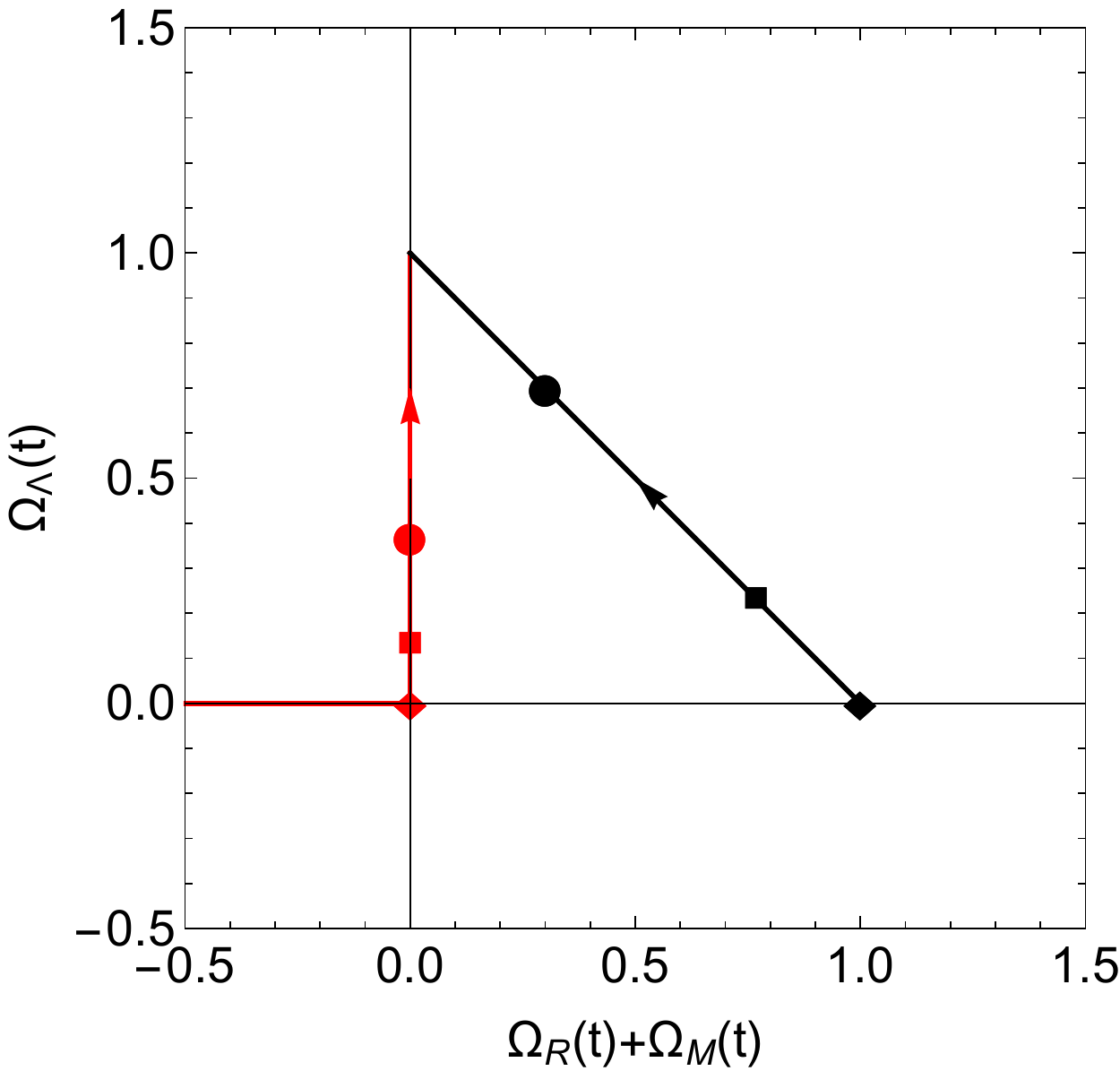}
  \caption{The dark energy contribution vs the sum of the radiation and matter contributions for $\Lambda$CDM in black and Mannheim in red. The dots denote values at $z=0$, the squares denote values at $z=1$, and the diamonds denote values at $z=10$.}
  \label{fig:Omega}
\end{figure}

In contrast to $\Lambda$CDM, which has a radiation era, then matter dominated era followed by the dark energy era, Mannheim's model has three phases: a quadratic expansion from the minimum radius in the early universe driven by radiation, a long period of linear growth via curvature, and an exponential era due to negative Higgs vacuum energy. Mannheim's cosmology always has positive acceleration, unlike $\Lambda$CDM which has a decelerating phase when matter or radiation are dominant. Figure \ref{fig:Omega} plots $\Omega_{\Lambda}(t)$ against $\Omega_{\text{R}}(t)+\Omega_{\text{M}}(t)$. $\Lambda$CDM begins close to $(1,0)$ and progresses along the flat-geometry diagonal line to $(0,1)$. However, this requires fine-tuning of the initial conditions, because a small deviation from $(0,1)$ leads to large excursions from flat geometry, before arriving at $(0,1)$. The Mannheim cosmology begins at $(-\infty,0)$ and evolves rapidly towards the origin, where it stays for 32 Gyr in the curvature-dominated era before moving up the y-axis towards $(0,1)$. CG predicts that the age of the Universe is approximately 32 Gyr, which is roughly 2.5 times greater than the $\Lambda$CDM age. For both the flat $\Lambda$CDM model and the Mannheim model, the point $(0,1)$ is an attractor. 
    
\section{Luminosity Distances}

\begin{table}
\begin{center}
  \begin{tabular}{ | c | c | c || c | c | c |}
    \hline
    $z$ & $\mu$ & $\pm$ & $z$ & $\mu$ & $\pm$ \\ \hline
    0.079 & 37.13 & 0.34 & 1.56 & 45.74 & 0.19 \\
    0.16 & 39.14 & 0.38 & 1.76 & 46.05 & 0.23 \\
    0.26 & 40.80 & 0.50 & 1.98 & 46.17 & 0.23 \\
    0.36 & 41.43 & 0.42 & 2.25 & 46.49 & 0.27 \\
    0.46 & 42.54 & 0.98 & 2.53 & 47.00 & 0.35 \\
    0.60 & 42.46 & 0.27 & 2.86 & 47.39 & 0.35 \\
    0.76 & 43.72 & 0.35 & 3.23 & 47.67 & 0.43 \\
    0.85 & 43.84 & 0.39 & 3.65 & 47.08 & 0.43 \\
    0.96 & 44.39 & 0.31 & 4.12 & 48.42 & 0.39 \\
    1.08 & 44.28 & 0.35 & 4.65 & 47.87 & 0.39 \\
    1.23 & 44.67 & 0.19 & 5.26 & 47.79 & 0.67 \\
    1.38 & 44.91 & 0.27 & 5.93 & 48.18 & 0.83 \\ \hline
  \end{tabular}
\end{center}
\caption{Binned quasar distance modulus data from \citet{Risaliti}. }
\label{TableQ}
\end{table}

\begin{table}
\begin{center}
  \begin{tabular}{ | c | c | c }
    \hline
    $\log(1+z)$ & $\mu$ & $\pm$ \\ \hline
    0.42 & 45.21 & 0.33 \\ 
    0.53 & 46.43 & 0.31 \\ 
    0.63 & 46.91 & 0.37 \\ 
    0.73 & 47.66 & 0.56 \\ 
    0.82 & 48.82 & 0.76 \\ 
    0.92 & 50.1 & 1.21 \\
    \hline
  \end{tabular}
\end{center}
\caption{GRB distance modulus data from \citet{Liu}, which we have binned into bins of 0.1 dex wide. The redshift quoted for each bin is the average redshift of all the GRBs within that bin.}
\label{TableGRB}
\end{table}

\begin{figure}
 \includegraphics[width=\columnwidth]{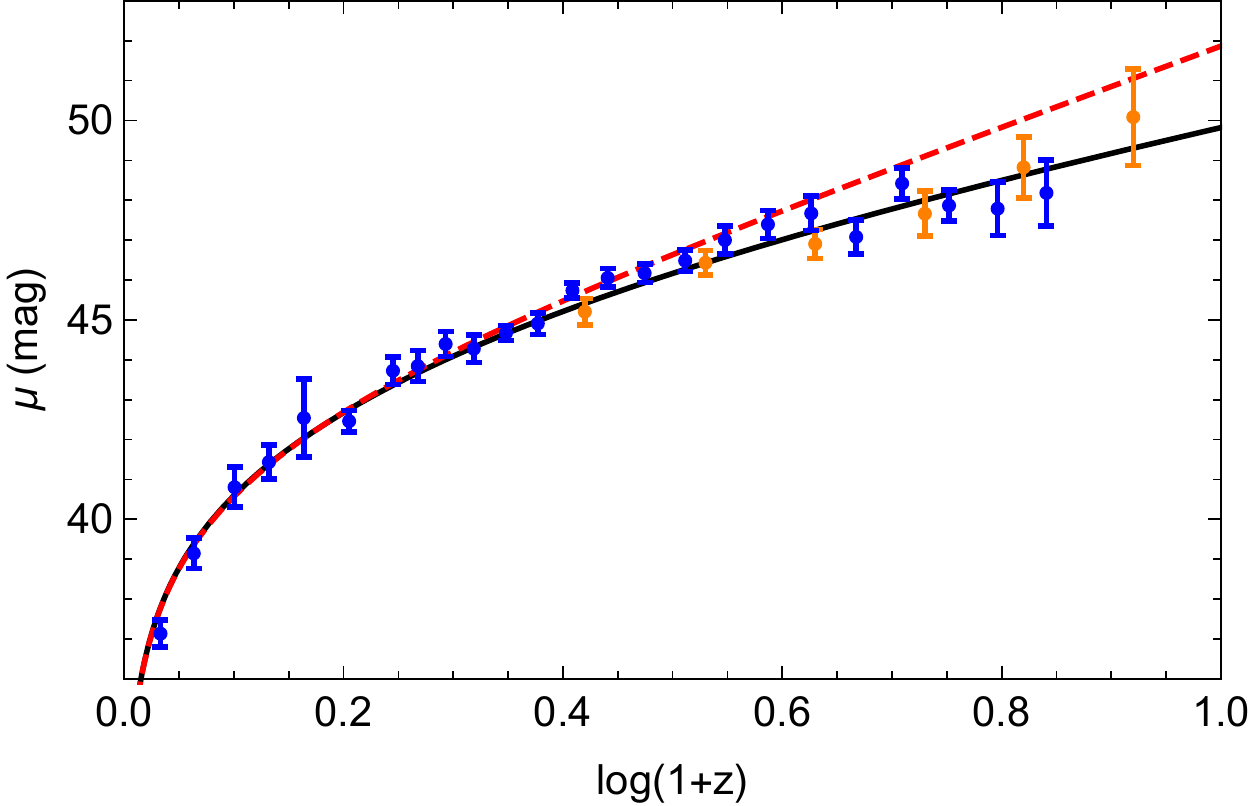}
 \caption{Here we present the Hubble diagram for both cosmologies: $\Lambda$CDM is the black line and Mannheim is the red dashed line. The binned GRB data are the orange points, the binned quasar data are in blue; each with their 1-$\sigma$ error bars. Mannheim's cosmology deviates significantly from the data and from the $\Lambda$CDM prediction at redshifts $z>2$.}
 \label{fig:HubbleDiagram}
\end{figure}

\citet{Mannheim2006} showed that $\Lambda$CDM and CG provide equally good fits to SNIa distances out to $z\sim$1. Here in Fig \ref{fig:HubbleDiagram}, we extend the Hubble diagram to $z\sim$8 by including GRB and quasar distance moduli. The distance modulus, $\mu$, is:
\begin{equation}
    \mu=25+5\log_{10}(D_{\text{L}}/\text{Mpc})\, ,
    \label{distmod}
\end{equation}
where $D_{\text{L}}$, the luminosity distance, is given by:
\begin{equation}
D_{\text{L}}=(1+z)\,\frac{c}{H_{0}}\frac{1}{\sqrt{|\Omega_{\text{K}}|}}\,S_{k}\left(\sqrt{|\Omega_{\text{K}}|}\,\int_{0}^{z}\frac{H_0}{H(z')}\,\mathrm{d}z'\right) \,,   
\end{equation}
with $S_{k}(x)=(\sin x,\, x,\, \sinh x)$ for $(k>0,\,k=0,\,k<0)$ respectively.

We use the binned dataset from Figure 5 of \citet{Risaliti}, comprising 808 quasars, in 24 redshift bins that are 0.1 dex in width, as shown in Table \ref{TableQ}. The quasars' distance moduli are calibrated using an empirical relationship between the UV and X-ray fluxes of quasars.

For the GRBs, we utilise the Mayflower sample of 79 GRBs from \citet{Liu}, shown in Table \ref{TableGRB}, in which they calibrate the Amati relation via the Pad\`{e} approximant method for 138 GRBs, with $z<1.4$, with the Union 2.1 SNIa data. We used 0.1 dex bins in $\log(1+z)$ to produce the averages in Table \ref{TableGRB}

With these distance moduli in hand, we produce the Hubble diagram for both these cosmologies in Figure \ref{fig:HubbleDiagram} using Eqn (\ref{distmod}) and the cosmological solutions for each theory: $\Lambda$CDM in Eqs (\ref{LambdaEqn}) and (\ref{Eqn6}) and Mannheim in Eqs (\ref{MannEqn}) and (\ref{Eqn13}). The quasar data points are blue and the GRB points are orange. Mannheim's cosmology is indistinguishable from $\Lambda$CDM for redshifts $z<1$ \citep{Mannheim1990}. However, at higher redshift, CG deviates greatly from $\Lambda$CDM. We perform the $\chi^2$ test for each cosmology, using the 79 unbinned GRB points and the 24 binned quasar points for a total of 103 data points, and find the following: $\chi_{\Lambda\text{CDM}}^2=62.4$; $\chi_{\text{Mannheim}}^2=111$. The flat $\Lambda$CDM and Mannheim models both have one free parameter. Thus, given $\Delta\chi^2=48$, we conclude that $\Lambda$CDM is by far the best fitting model of either cosmology examined.

\section{Discussion}

In this section, we examine some of the theoretical problems faced by $\Lambda$CDM and CG cosmologies. One problem for the $\Lambda$CDM cosmology is the cosmological constant problem \citep{Weinberg1989}, or $\Lambda$ fine-tuning problem, where the predicted value from standard particle physics results in an extremely high dark energy component, $\Omega_{\Lambda}\sim10^{120}$ if it is associated to the Planck scale, $\Omega_{\Lambda}\sim10^{60}$ if the electroweak scale. CG cosmology has a self-quenching mechanism \citep{Mannheim2006}, so that for any vacuum energy $\lambda S_0^4<0$ (where $S_0$ is the value of the Higgs field in the Higgs frame), $\Omega_{\Lambda}(t)$ must lie within the range $0\leq\Omega_{\Lambda}(t)\leq1$; see Figure \ref{fig:Omega}. 

However, we have identified a similar issue in Mannheim's cosmology: the $\lambda$ fine-tuning problem. Similar to the $\Lambda$ fine-tuning problem in $\Lambda$CDM, we find that to fit Mannheim's preferred parameters the Higgs self-coupling constant must be tiny, on the order of $-10^{-176}$. Recalling the definition of $\bar{\Omega}_{\Lambda},\, \rho_{\Lambda},$ and $\epsilon$, we may rewrite $\bar{\Omega}_{\Lambda}$ as:
\begin{equation}
    \bar{\Omega}_{\Lambda}=-\frac{2\lambda S_0^2}{H_0^2}\, .
\end{equation}

Now by rearranging the above equation for $\lambda$ and by using $S_0^2=\frac{3}{4\pi\epsilon G}=\frac{2\rho_c}{\epsilon H_0^2}$ yields:
\begin{equation}
    \lambda=\frac{-\epsilon\,\bar{\Omega}_{\Lambda}\,H_0^4}{4\rho_c}\,.
    \label{lambda1}
\end{equation}

By defining the radiation energy-density as $\rho_R=4\sigma T_{\rm{CMB}}^4$, where $T_{\rm{CMB}}$ is the present temperature of the CMB, we can rearrange the definition of $\bar{\Omega}_{\rm{R}}$ to get:
\begin{equation}
    \epsilon=\frac{-\bar{\Omega}_{\rm{R}}\,\rho_c}{4\sigma\, T_{\rm{CMB}}^4}\,.
    \label{epsilon}
\end{equation}

Then by substituting Eqn (\ref{epsilon}) into Eqn (\ref{lambda1}) and returning to SI units, we find that:
\begin{equation}
    \lambda=\frac{\hbar}{c^2}\frac{\bar{\Omega}_{\Lambda}\,\bar{\Omega}_{\rm{R}}\,H_0^4}{16\sigma\, T_{\rm{CMB}}^4}\,.
\end{equation}

The supernova distances give $\bar{\Omega}_{\Lambda}=0.37$, and Mannheim adopts $\bar{\Omega}_{\rm{R}}\approx-10^{-60}$ so that $T_{\rm{max}}=T_{\rm{CMB}}/a_{\rm{min}}$ exceeds the electroweak scale. Using the observed values of $H_0$ and $T_{\rm{CMB}}$, we find that Mannheim's model predicts a value for $\lambda$ that is of the order $\lambda \approx -10^{-176}$. This is very small, why is it not zero? One could appeal to some of the same proposed solutions to the fine-tuning problem in $\Lambda$CDM here, such as the existence of a multiverse.

$\Lambda$CDM also has the cosmic coincidence problem, where a high degree of fine tuning is required so that $\Omega_{\Lambda}\sim\Omega_{\text{M}}$ at current time. \citet{Mannheim2006} claims that his model solves the cosmic coincidence problem, because $\Omega_{M}\ll \Omega_{\Lambda}$. However, there remains a similar cosmic coincidence problem; that the current era just so happens to be when the Universe transitions from curvature to dark energy driven expansion, or that $\Omega_{\Lambda}\sim\Omega_{\text{K}}$ which also requires fine tuning. Thus CG and $\Lambda$CDM have similar fine-tuning problems.

\section{Conclusions}

We have compared Mannheim's CG cosmology to the standard $\Lambda$CDM cosmology. \citet{Mannheim2006} indicated close agreement with $\Lambda$CDM for the predicted distance moduli for redshift $z<1$. By collating GRB \citep{Liu} and quasar \citep{Risaliti} data, we have extended the Hubble diagram out to $z=8$ and reanalysed the predictions of the CG cosmology. We find that the CG cosmology deviates significantly from $\Lambda$CDM and the data for redshift $z>2$. We performed the $\chi^2$ test and found that $\Lambda$CDM is the favoured model with $\Delta\chi^2=48$.

We have discussed theoretical problems with these cosmological models. \citet{Mannheim2006} claims that his model solves the cosmic coincidence problem. However, in this paper we found an analogue to the cosmic coincidence problem, involving $\Omega_{\rm{K}}$ and $\Omega_{\Lambda}$, for Mannheim's model. Additionally, we identified a $\lambda$ fine-tuning problem analogous to that of $\Lambda$CDM. We determined that using Mannheim's model, the Higgs self-coupling constant is of the order $\lambda\approx-10^{-176}$. In summary, $\Lambda$CDM and CG have similar fine-tuning issues, but $\Lambda$CDM fits the data far better than CG.

\section*{Acknowledgements}
CR is supported by STFC studentship ST/N504427/1. KH acknowledges support from STFC Consolidated grant ST/M001296/1




\bibliographystyle{mnras}
\bibliography{canon} 



\appendix

\section{Derivation of Mannheim's Cosmology}

This treatment follows that of \citet{Mannheim2006}. The combined action is $S_{\rm{total}}=S_{\rm{Weyl}}+S_{M}$, where the matter action is:
\begin{equation}
\begin{split}
    S_{M}=-\int d^{4}x\, \sqrt{-g}\, \left( \frac{1}{2}[S^{;\mu}S_{;\mu}-\frac{1}{6}S^{2}R] +\lambda S^{4} +\bar{\psi}(\slashed{D}-hS)\psi\right) \ ,
    \end{split}
\end{equation}
where $S(x)$ is the Higgs scalar field; $R$ is the Ricci scalar; $\lambda$ is the Higgs self-coupling constant; $\psi(x)$ is the fermion's wave function; $h$ is the Yukawa coupling constant between the Higgs field and the fermions;  $\slashed{D}\equiv i\gamma^{\mu}D_{\mu}$ is the slashed Dirac operator; and $D_{\mu}\equiv \partial_{\mu}+\Gamma_{\mu}$ is the Dirac operator, where $\Gamma^{\mu}(x)$ is the fermion spin connection and $\gamma^{\mu}(x)$ are the gamma matrices.

The contribution $S^{;\mu}S_{;\mu}-\frac{1}{6}S^{2}\,R$ is a conformally invariant addition to the matter action. Now, we take the variation with respect to $\psi$, $S$ and the metric to obtain the following equations of motion.

The Dirac equation:
\begin{equation}
    (\slashed{D}-hS)\psi=0\ .
    \label{B2}
\end{equation}

The Higgs equation:
\begin{equation}
    S^{;\mu}_{;\mu}+\frac{R}{6}S-4\lambda S^{3}+h\bar{\psi}\psi=0\ .
    \label{B3}
\end{equation}

The conformal stress-energy tensor:
\begin{equation}
\begin{split}
    T_{\mu\nu}=i\bar{\psi}\gamma_{\mu}D_{\nu}\psi+\frac{2}{3}S_{;\mu}S_{;\nu}-\frac{1}{6}g_{\mu\nu}S^{;\alpha}_{;\alpha} \\ -\frac{1}{3}SS_{;\mu;\nu}+\frac{1}{3}g_{\mu\nu}SS^{;\alpha}_{;\alpha}-\frac{1}{6}S^{2}\left(R_{\mu\nu}-\frac{1}{2}g_{\mu\nu}R\right) \\
    -g_{\mu\nu}\left(\lambda S^{4}+\bar{\psi}(\slashed{D}-hS)\psi \right)\ .
    \label{B4}
\end{split}    
\end{equation}

Inserting (\ref{B2}) into (\ref{B4}) yields:
\begin{equation}
\begin{split}
    T_{\mu\nu}=i\bar{\psi}\gamma_{\mu}D_{\nu}\psi+\frac{2}{3}S_{;\mu}S_{;\nu}-\frac{1}{6}g_{\mu\nu}S^{;\alpha}_{;\alpha} \\ -\frac{1}{3}SS_{;\mu;\nu}+\frac{1}{3}g_{\mu\nu}SS^{;\alpha}_{;\alpha}-\frac{1}{6}S^{2}\left(R_{\mu\nu}-\frac{1}{2}g_{\mu\nu}R\right) \\
    -\lambda S^{4}g_{\mu\nu}\ .
    \label{B5}
\end{split}    
\end{equation}

\citet{Mannheim2006} uses a conformal transformation to the Higgs frame, where $S=S_0$, so that fermion mass is a space-time constant, $hS\rightarrow hS_0$. (\ref{B5}) becomes:
\begin{equation}
    T_{\mu\nu}=i\bar{\psi}\gamma_{\mu}D_{\nu}\psi-\frac{1}{6}S_{0}^{2}\,G_{\mu\nu}-g_{\mu\nu}\lambda S_{0}^{4}\ .
    \label{Import}
\end{equation}

The incoherent averaging of $i\bar{\psi}\gamma_{\mu}D_{\nu}\psi$ \citep{Mannheim1990} allows a perfect fluid representation for the ordinary matter and radiation:  $(\rho+p)U_{\mu}U_{\nu}+pg_{\mu\nu}$ with pressure $p$, energy density $\rho$ and $U_{\alpha}$ is the fluid's 4-velocity. Hence, (\ref{Import}) is now:

\begin{equation}
    T_{\mu\nu}=(\rho+p)U_{\mu}U_{\nu}+(p-\rho_{\Lambda})g_{\mu\nu}-\frac{1}{6}S_{0}^{2}\,G_{\mu\nu} \ ,
    \label{Export}
\end{equation}
where $\rho_{\Lambda}=\lambda S_{0}^{4}$ is the Higgs vacuum energy density.
The equivalent equation to the Einstein equation of CG is the Bach equation:

\begin{equation}
    4\alpha_{g} W_{\mu\nu}=T_{\mu\nu}\ ,
    \label{A8}
\end{equation}
where $\alpha_{g}$ is a dimensionless constant and  $W_{\mu\nu}$ is the Bach tensor, created by varying the Weyl action with respect to the metric. The FRW metric is a conformally flat metric, thus the Weyl tensor, and hence the Bach tensor, vanishes. Thus (\ref{A8}) becomes:
\begin{equation}
    0=T_{\mu\nu}\ .
\end{equation}

Thus, to obtain Mannheim's cosmology, rearrange Equation (\ref{Export}) to find:
\begin{equation}
    G_{\mu\nu}=8\pi\epsilon G\left((\rho+p)U_{\mu}U_{\nu}+(p-\rho_{\Lambda})g_{\mu\nu}\right)\ ,
    \label{eqn:M1}
\end{equation}
where the effective gravitational constant is $G_{\rm{eff}}=-\epsilon G$ with $\epsilon\equiv\ \frac{3}{4\pi S_{0}^{2}G}$.


\bsp	
\label{lastpage}
\end{document}